# Enhanced biochemical sensing with high-Q transmission resonances in free-standing membrane metasurfaces

Samir Rosas[1], Wihan Adi[1], Aidana Beisenova[1], Shovasis Kumar Biswas[2], Furkan Kuruoglu[1,3], Hongyan Mei[2], Mikhail A. Kats[2], David A. Czaplewski[4], Yuri S. Kivshar[5], Filiz Yesilkoy[1]*

[1] Department of Biomedical Engineering, University of Wisconsin-Madison Madison, WI 53706, USA

[2] Department of Electrical and Computer Engineering, University of Wisconsin-Madison Madison, WI 53706, USA

[3] Department of Physics, Faculty of Science, Istanbul University, Vezneciler, 34134, Istanbul, Turkey

[4]Center for Nanoscale Materials, Argonne National Laboratory, Lemont, IL, 60439, USA

[5]Nonlinear Physics Center, Australian National University, Canberra ACT 2601, Australia

*Corresponding author. Email: filiz.yesilkoy@wisc.edu

## Abstract

Optical metasurfaces provide novel solutions to label-free biochemical sensing by localizing light resonantly beyond the diffraction limit, thereby selectively enhancing light-matter interactions for improved analytical performance. However, high-Q resonances in metasurfaces are usually achieved in the reflection mode, which impedes metasurface integration into compact imaging systems. Here, we demonstrate a novel metasurface platform for advanced biochemical sensing based on the physics of the bound states in the continuum (BIC) and electromagnetically induced transparency (EIT) modes, which arise when two interfering resonances from a periodic pattern of tilted elliptic holes overlap both spectrally and spatially, creating *a narrow transparency window* in the mid-infrared spectrum. We experimentally measure these resonant peaks observed in transmission mode (Q~734 @ $\lambda$~8.8 $\mu m$) in free-standing silicon membranes and confirm their tunability through geometric scaling. We also demonstrate the strong coupling of the BIC-EIT modes with a thinly coated PMMA film on the metasurface, characterized by a large Rabi splitting (32 cm$^{-1}$) and biosensing of protein monolayers in transmission mode. Our new photonic platform can facilitate the integration of metasurface biochemical sensors into compact and monolithic optical systems while being compatible with scalable manufacturing, thereby clearing the way for on-site biochemical sensing in everyday applications.

# INTRODUCTION

Optical metasurfaces are ultrathin engineered interfaces with periodic subwavelength structures forming unit cells designed to efficiently modulate the fundamental properties of light. An important function of resonant metasurfaces is their capability to localize the incident electromagnetic field into sub-wavelength volumes, where light-matter interactions are enhanced(*1*, *2*). The efficiency of near-field light-matter coupling strongly depends on the quality (Q) factor of the metasurface resonance, which is related to the photon lifetime in a cavity(*3*, *4*). Dielectric metasurfaces made from high-index and low-loss materials can realize significantly sharp resonances compared to their plasmonic counterparts, where the intrinsic metallic losses limit their Q-factors. Because dielectric metasurfaces supporting quasi-bound states in the continuum (q-BIC) modes enable precise control over radiation channels, they have been integrated into a variety of optical applications, including lasing (*5*), optical trapping (*6*, *7*), nonlinear and quantum optics (*8*, *9*). Specifically in the mid-IR spectrum ($\lambda$ =2.5–25 µm), multi-resonant q-BIC metasurface arrays facilitated imaging-based surface-enhanced infrared absorption spectroscopy (SEIRAS) with high spectral resolution (*10*), frequency conversion processes (*11*), and vibrational strong coupling (*12*, *13*). Despite encouraging progress, challenges related to integration into miniaturized robust optical systems and low-cost scalable manufacturing have, thus far, impeded mid-IR metasurfaces from reaching their full potential and addressing the unmet needs of real-world biochemical sensing applications.

Conventional mid-IR metasurfaces supporting q-BIC resonances have been primarily fabricated on IR-transparent supporting substrates by patterning an array of sub-wavelength structures made from low-loss and high-index materials, such as Si (*10*) and Ge (*14*). In this approach, the need for the supporting substrates is a critical bottleneck for the mass-scalable manufacturing of the mid-IR metasurfaces because transparent materials over the mid-IR window, e.g., $CaF_2$ and $MgF_2$, are not compatible with wafer-scale high-throughput fabrication processes. To eliminate the need for IR-transparent supporting substrates, we recently proposed patterning subwavelength periodic voids into free-standing Si membrane metasurfaces and demonstrated high-Q Brillion zone folding (BZF)-induced q-BIC modes(*13*). Others also reported fabricating mid-IR nanoantennas on silicon nitride membranes (*15*) and metasurfaces patterned into silicon carbide (SiC) membranes to eliminate substrate effects (*16*). While these are promising developments toward scalable fabrication of mid-IR metasurfaces, challenges related to their integration into compact instruments remain.

The key approach in implementing high-Q dielectric metasurfaces for SEIRAS-based biochemical sensing is that the q-BIC resonances supported by these metasurfaces generate resonance dips in transmission or peaks in reflection (*17*). The fundamental detection signal in SEIRAS is usually a perturbation in frequency, amplitude, and Q factor of the resonance, induced by the enhanced coupling between the vibrational modes of molecules and metasurface resonances, enabling the extraction of the molecular fingerprints. For example, when working with resonances characterized by spectral dips, the spectral resonance parameters must be extracted in low-light conditions, leading to low signal-to-noise ratios (SNRs) and longer integration times. Therefore, reflection mode measurements, yielding spectral peaks in the far field, have been frequently used to interrogate the sharp q-BIC modes in SEIRAS applications (*10*, *14*, *17*). However, reflection measurements require complex optical paths, with extra optical components, such as beam splitters and mirrors, hindering the metasurface device integration into robust collinear imaging platforms and the development of compact on-site biochemical sensors.

Here, we suggest a completely different approach to SEIRAS by designing and implementing high-Q resonances in the transmission mode. We reveal that such resonances can be supported by free-standing membrane metasurfaces with tilted rod apertures, as shown in Fig 1. This feature makes the membrane metasurfaces well-suited for *three key functionalities*: optical filtering, strong coupling, and biochemical sensing. Our resonance mechanism is driven by a metasurface analogue of electromagnetically induced transparency (EIT) (*18–21*), where a broad multipolar surface lattice mode (SLM) comprising elements of an electric dipole and a magnetic quadrupole interferes with an ultra-sharp q-BIC mode induced by breaking the in-plane symmetry in a tilted rod pair design (Fig. 1a). When the two interfering resonances generated by a periodic void pattern carved into a 1μm thick Si membrane spectrally and spatially overlap, a narrow transparency window emerges in the mid-IR spectrum (Fig. 1b). Previously, metasurface analogues of EIT modes have been demonstrated by interfering plasmonic resonances (*20*) and dielectric nanoantennas for refractometric sensing applications in the near-IR (*18*, *21*). In this study, for the first time, we experimentally present a high-Q (~734 @ $\lambda$~8.8 μm) EIT-like mode in the mid-IR leveraging the unique strengths of the q-BIC resonances and all-dielectric membrane metasurfaces. In addition, we show the Q-factor and frequency tunability of these resonances via geometric design parameters. Finally, we report experimental results investigating light-matter interactions in these powerful metasurface resonances. First, we show vibrational strong coupling with a thinly coated PMMA film on the metasurface, characterized by significantly large Rabi splitting (32 $cm^{-1}$). Second, we demonstrate biosensing of protein monolayers in transmission mode, enabled by the SEIRAS (Fig. 1c). Overall, our approach introduces elegant high-Q transmissive resonances that are ideal for compact metadevices while being compatible with scalable manufacturing, clearing the way towards on-site biochemical sensing in everyday applications.

## RESULTS

### Design and fabrication of membrane metasurfaces

Our metasurfaces consist of free-standing, intrinsic, single crystal ⟨100⟩ silicon membranes of thickness h=1 μm patterned with an array of unit cells consisting of pairs of tilted elliptical rods with mirror symmetry across the major vertical axis, illustrated in Fig. 1a. These rods are *etched through the membrane* and have minor and major axis *a*, and *b*, respectively. The geometric parameters of a single unit cell are detailed in Fig. 3a. The tilting angle $\theta$ is measured with respect to the vertical axis, and unit cells are arranged in a rectangular lattice with periods $P_x$, and $P_y$. In optimizing the metasurface design parameters, we aimed to achieve the following criteria: (1) the metasurface must have a resonance peak in transmission with an amplitude approaching unity when illuminated with normally incident light; (2) high electromagnetic-field confinement into accessible photonic cavities at resonance; (3) easy tunability of the transmittance peak via geometric design parameters to allow for operation in the broad mid-infrared fingerprint spectral range; (4) a high and tunable Q-factor of the resonance peak; and (5) a straightforward fabrication process compatible with mass manufacturing. We performed numerical simulations using commercially available software packages running finite-element frequency-domain (FEFD) and finite-difference time-domain (FDTD) solvers (see Methods). We identified the following metasurface design parameters satisfying the above criteria: a = 1.4 µm, b = 4.0 µm, $P_x$ = 4.0 µm, $P_y$ = 4.4 µm, h = 1.0 µm (Fig. 3a). We tuned the resonance peak position by scaling the geometric parameters of the unit cell with the factor (S) and fabricated metasurfaces with tilting angle from $\theta = 0°$ to 10° to adjust Q factor.

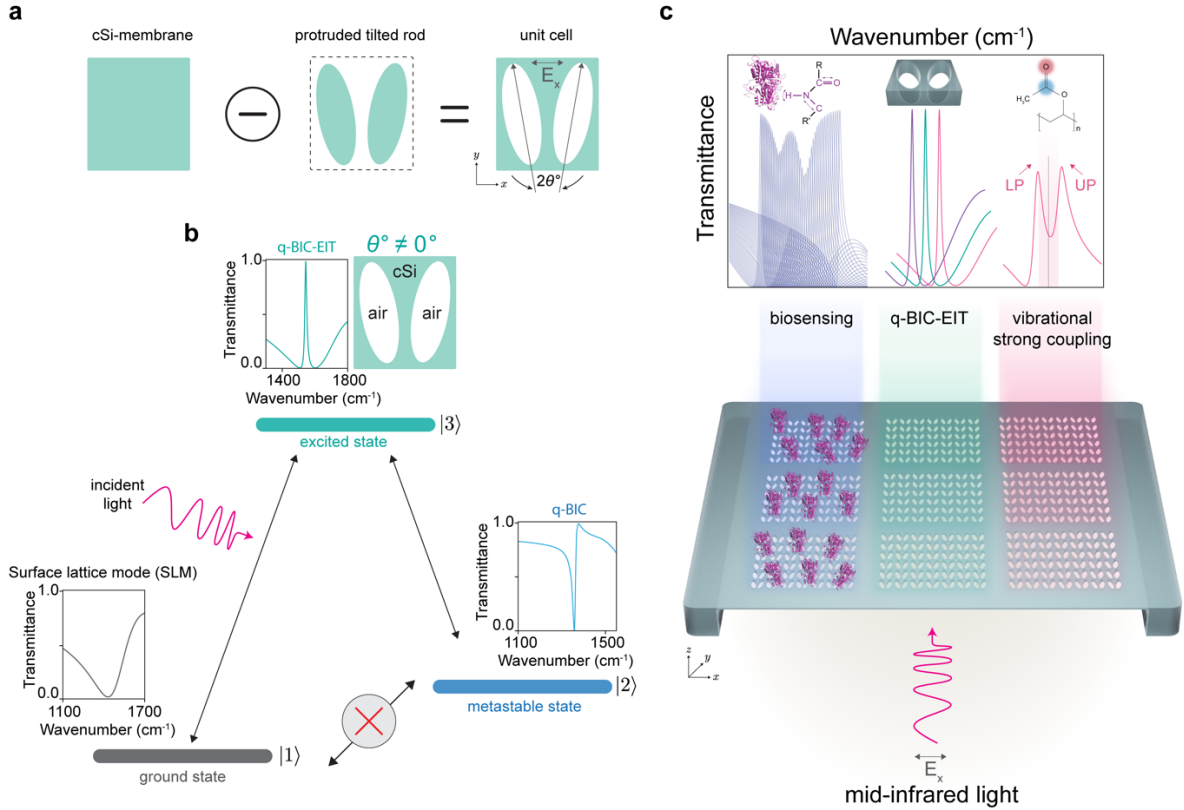

**Figure 1. Free-standing membrane metasurfaces with transmissive resonances for mid-infrared biochemical sensing applications on a collinear optical imaging setup.**
**a**, Illustration of the metasurface unit cell including a pair of elliptical rods etched through a crystalline Silicon (cSi) free-standing membrane slab. Like standard protruded resonators, when the rod voids are tilted oppositely with respect to each other along their long axes, the BIC modes couple to continuum, turning into the q-BIC modes. **b**, Analogous to a three-level atomic ensemble, the photonic analogue of the EIT mode is modeled with Λ-shaped energy levels, including a ground state $|1\rangle$, an excited state $|3\rangle$, and a metastable state $|2\rangle$. In our metasurface design, the transition $|1\rangle-|3\rangle$ is a surface lattice mode (SLM) with a high decay rate ($\Gamma_{SLM}$), which can be directly excited with a linearly polarized light along the x-axis ($E_x$). The transition $|2\rangle-|3\rangle$ is characterized by a BIC mode, whose decay rate ($\Gamma_{BIC}$) is near zero ($\Gamma_{SLM} \gg \Gamma_{BIC}$). Since, ideal BIC ($\theta = 0°$) does not directly couple to the free-space propagation, here we represent it with a q-BIC by introducing a small tilting angle to the rods ($\theta \neq 0°$). In the presence of an incident x-polarized light at normal incidence, a quantum interference occurs, and the transition amplitudes of $|1\rangle-|3\rangle$ and $|2\rangle-|3\rangle$ destructively interfere, creating a narrow transparency window in the spectrum, which we call q-BIC-EIT. **c**, Free-standing membrane metasurfaces supporting q-BIC-EIT modes are used to detect protein monolayers via SEIRAS and enables vibrational strong coupling at room temperature.

The metasurfaces were fabricated by patterning free-standing cSi membranes using electron-beam lithography (EBL) for mask formation and reactive-ion etching (RIE) to anisotropically etch Si, generating punch-through holes with vertical and smooth sidewalls (see Methods). A brightfield photograph of a fabricated Bucky-metasurface, composed of hundreds of thousands of tilted rod voids, is shown in Fig. 2a. In this image, the dark regions are unpatterned, and the colorful regions are patterned membrane metasurface areas. Scanning electron microscope (SEM) images of a fabricated metasurface are shown in Fig. 2b, with an angled view of the tilted rod voids in the inset, showing vertical sidewalls through the 1 µm thick membrane.

**Metasurface analogue of electromagnetically induced transparency (EIT)**

The experimentally measured transmittance in Fig. 2c was obtained by illuminating the sample with collimated, normally incident, x-polarized, mid-infrared light. The metasurface exhibits an EIT-like peak in transmission, with ~80% transmittance in the mid-infrared spectral range. For comparison, transmittance through the unpatterned membrane region is also plotted in Fig. 2c. When we imaged the Bucky-metasurface at the peak of its resonance wavenumber (1458 cm$^{-1}$), the patterned metasurface regions light up with high contrast relative to the background, clearly delineating the Bucky pattern (Fig. 2d). When illuminated at 1502 cm$^{-1}$, where metasurface suppresses the transmission more than the unpatterned membrane, the inverse Bucky pattern is captured. Conversely, when the spectrum from the patterned region intersects with the spectrum of the unpatterned regions (1478 cm$^{-1}$), the overall membrane displays an almost uniform transmission, generating a low-contrast image of Bucky. Interestingly, in this image, the boundaries between the patterned and unpatterned regions exhibit the highest contrast due to metasurface-size-dependent resonance intensity changes at the narrow regions of metasurfaces.

Our metasurface design supports two distinct modes that overlap in space and frequency but with different resonance mode linewidths, as illustrated in Figure 1b. The first mode, a surface lattice mode (SLM), originates from the enhanced radiative coupling of localized Mie resonances associated with individual tilted rod voids embedded in the free-standing cSi membrane. These collective resonances emerge from Rayleigh anomalies, where enhanced radiative coupling is due to in-plane diffraction orders of the lattice (*21*, *22*). The SLMs can be tuned by varying the lattice period, $P_x$, or $P_y$. The second mode is a q-BIC mode and belongs to a unique class of wave phenomena where localized waves remain confined despite coexisting with a continuous spectrum of radiating waves that can carry energy away. Of the many ways to excite q-BICs, in-plane symmetry breaking is one of the most popular and straightforward. By introducing controlled imperfections or asymmetries, ideal BICs can be converted into q-BICs with finite but high Q-factors (*23*, *24*). In our case, the symmetry breaking was introduced along the x-axis by setting the tilting angle $\theta \neq 0°$. Destructive interference between the SLM and q-BIC modes induces a narrow-band transparency window (Fig. 1b).

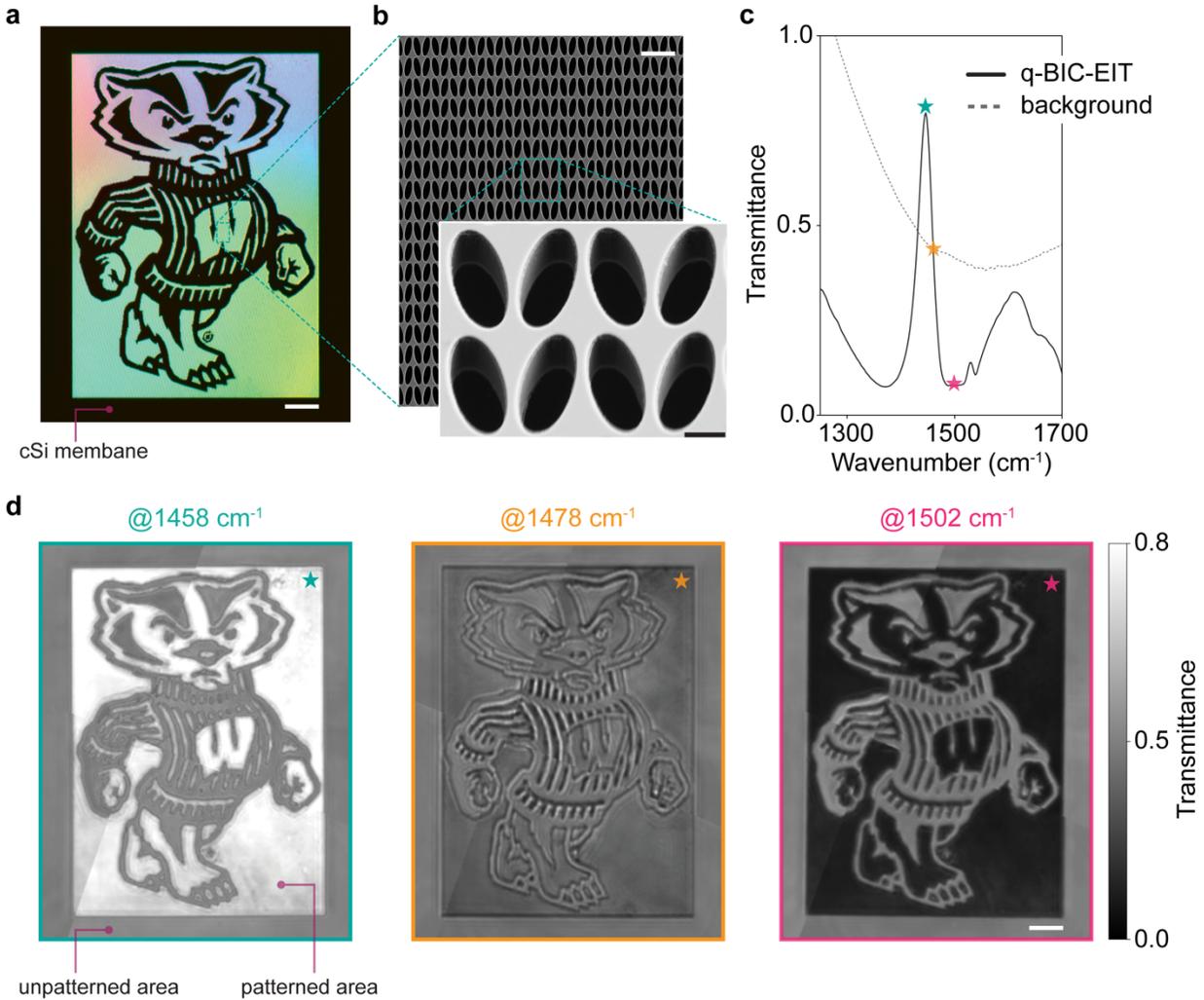

**Figure 2. Realization of free-standing cSi-membrane metasurfaces**. **a**, Brightfield photograph of the free-standing membrane bucky-metasurface composed of hundreds of thousands of tilted rod voids (scale bar 200 μm). **b**, Scanning electron micrograph (SEM) of the metasurface showing the zig-zag rod design (scale bar 8 μm) and cavities with vertical sidewalls on a 1 μm-thick cSi membrane of tilted rods (inset scale bar 1 μm). **c**, Experimentally measured q-BIC-EIT-like resonance transmittance spectrum of the free-standing metasurface and background, taken from patterned and unpatterned cSi membrane regions, respectively, with $\theta = 10°$. **d**, Infrared transmission microscopy image at the resonance peak (1458 cm$^{-1}$), at the wavenumber 1478 cm$^{-1}$, where transmittance through the metasurface and the background are similar, and at minimum transmittance (1502 cm$^{-1}$) of the metasurface. Note how the contrast of the bucky-metasurface microscopy images changes when transitioning from its highest transmittance value (~0.8) to its minimum (~0.065).

To understand the emergence of high-Q transmissive modes in our system, we studied the nature of the SLM and q-BIC resonances by examining the electric and magnetic near-field intensity enhancements relative to the incident field intensity at two distinct angles (Fig. 3). At $\theta = 0°$, the SLM couples to the incident light, resulting in the electric and magnetic field distributions as shown in Figure 3b. The field maps are the xy-plane cross-sections of the membrane at the center height ($z = h/2$) for illumination at 1460 cm$^{-1}$. Furthermore, we calculated the scattering cross-section ($\sigma_{scs}$) of cartesian multipoles in the far field to determine the multipolar decomposition of the SLM (the mathematical treatment is presented in Methods). Figure 3e shows that at $\theta = 0°$, the electric dipole (ED), and the magnetic quadrupole (MQ) are the main contributors to the SLM optical response, which agree with the field maps in Figure 3b.

When the tilting angle $\theta$ deviates from zero, the coupling between the SLM and q-BIC resonances leads to an increase in the electromagnetic field enhancements (see Fig. 3c and 3d), so the fields leak into the air media inside the tilted rods, generating accessible photonic cavities. The electric field's enhancement decays to 36% of its maximum value at a distance of 800 nm from the vertical walls of the tilted rods (see Fig. 3c and SI Fig. S2), with an effective mode volume of the unit-cell $V_{eff} = 0.065(\lambda/n_{eff})^3$ at the resonance (see Methods). The multipolar mode decomposition analysis revealed that the electric quadrupole (EQ) and magnetic dipole (MD) are the primary drivers of the transparency window resonance, with minimal contribution from the electric dipole (ED) and magnetic quadrupole (MQ) (see Fig. 3f), confirming the destructive nature of the SLM and q-BIC resonances. In addition, we show that the reconstructed transmission spectrum obtained from the multipolar decomposition analysis is in excellent agreement with the numerical simulations and experimental measurements (see Fig. S1).

To bring quantitative insights into the SLM and q-BIC resonances, we employed coupled-mode theory (CMT) to mathematically model the interplay between the SLM and q-BIC resonance modes (see SI and Fig. S1). We adopted the same approach as in ref. (*25*) to analytically understand the interaction between the low-Q SLM and the high-Q q-BIC modes with zero detuning in their resonance frequencies, analogous to the EIT effect observed in a three-level atom. In EIT-like resonances, the occurrence of the transparency window results from strong Fano interferences controlled by the coupling strength ($\kappa$) between the interacting modes. Under the right circumstances, when the threshold coupling strength ($\kappa_T$) satisfies the condition for which $\kappa < \kappa_T$ (see SI for a detailed derivation), the two resonances will undergo a transition into a weak driving regime. In this regime, spatially overlapping modes with identical resonance frequencies but different linewidths ($\Gamma_{SLM} \gg \Gamma_{BIC}$) result in a narrow transparency window, similar to EIT. The transmission in this regime is described by $T_{EIT} = 1 - 2\Gamma_R\chi_i$, where $\Gamma_R$ captures the radiative loss of the SLM mode and $\chi_i$ is the imaginary part of $\chi$ as defined in SI Eq. S4. Values for all $T_{EIT}$ parameters that could not be precisely determined were obtained by curve-fitting to the measured and simulated transmission spectra as shown in supplementary Fig. 1S, and Table S1. Not only does CMT explain the behavior of the coupling between the SML and the q-BIC modes, but it also aligns well with both the simulated and measured spectra.

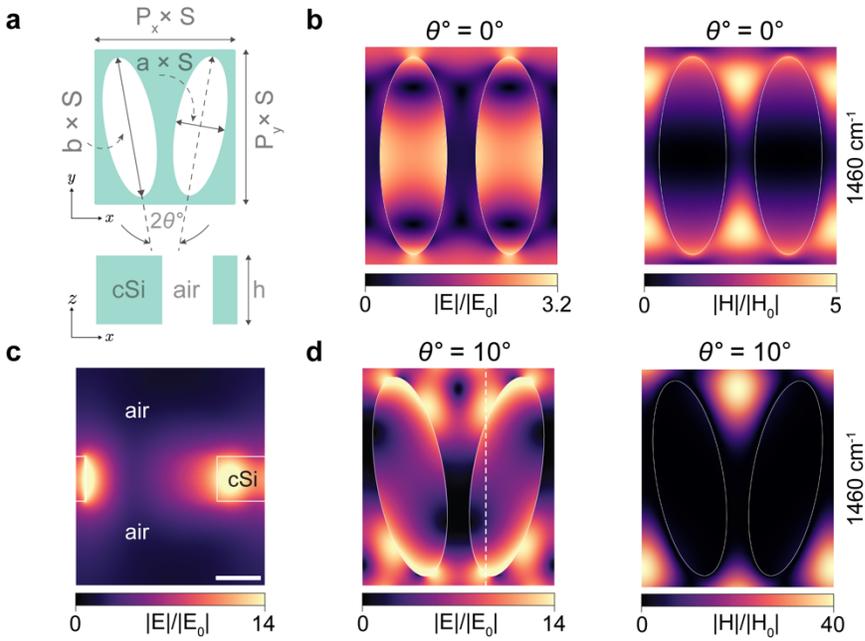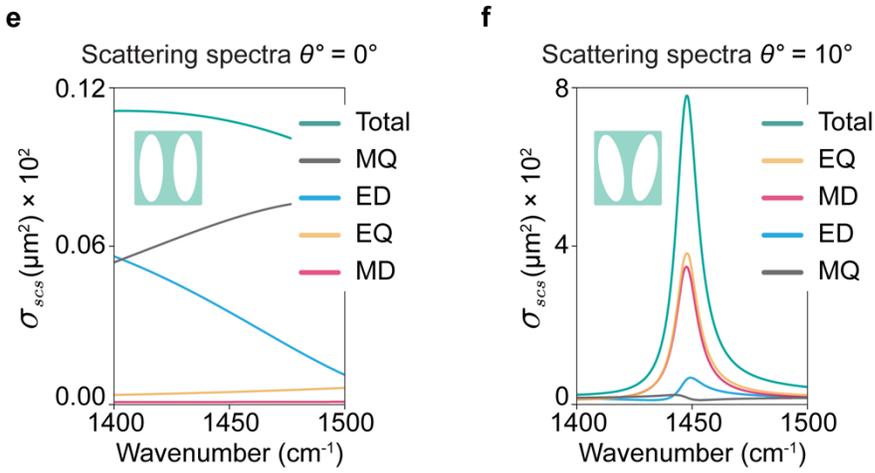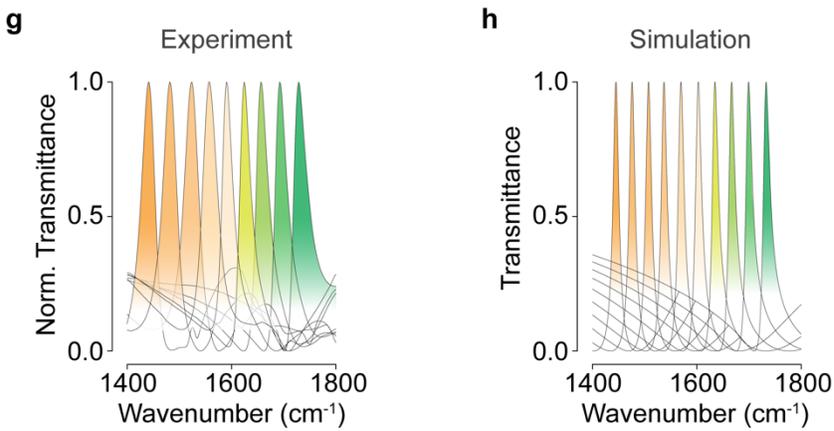

**Figure 3. Photonic resonance characteristics of the membrane metasurfaces. a**, The relevant geometrical design parameters for the unit cell in the illustration are a = 1.4 µm, b = 4.0 µm, $P_x$ = 4.0 µm, $P_y$ = 4.4 µm, h = 1.0 µm, S = 1, and $\theta$ = 10°. Electric and magnetic field maps of the unit cell at the middle depth plane (h/2) when **b,** $\theta = 0°$, and **d,** $\theta = 10°$ at 1460 cm$^{-1}$. Notice how both the electric and magnetic fields are enhanced by the symmetry breaking along the x-axis introduced by the tilting angle $\theta$. The electric field, in particular, reaches its maximum at the tips of the tilted rod voids. **c,** A longitudinal electric field profile, cut through the plane of the white dashed line (shown in **d**), demonstrates that the electric field decays to 36% of its maximum amplitude in the airholes at approximately 800 nm away from the vertical walls of the rod voids. The calculated effective mode volume at resonance peak ($\lambda_{res}$ = 1460 cm$^{-1}$) is $V_{eff} = 0.065(\lambda/n_{eff})^3$ with $n_{eff}$ = 2.21 (scale bar 1 µm). Multipole decomposition for **e,** $\theta$ = 0° and **f,** $\theta$ = 10°. At $\theta$ = 0°, the electric-dipole (ED) and the magnetic quadruple (MQ) are the main contributors to the scattering cross section of the SLM. In contrast, at $\theta$ = 10°, the electric-quadrupole (EQ) and the magnetic dipole (MD) drive the q-BIC-EIT transparency window. **g,** Experimentally measured and **h,** simulated transmittance spectra for a tilted rod metasurface with $\theta$ = 10°, where spectral resonance tunning was achieved by varying the scaling factor S of the metasurface unit cell from 0.7 to 1.0.

Moreover, the occurrence of q-BIC-EIT resonances is accompanied by the slow light effect. We calculate the group delay as well as the group index at the peak of the q-BIC-EIT resonance (~1460 cm$^{-1}$). The group delay is calculated using the formula $\tau_g = -d\phi_{EIT}/d\omega$, where $\phi_{EIT}$ is the transmission phase shift in radians, and $\omega$ the angular frequency. The group index is defined as $n_g = c/v_g = c\tau_g/h$ with $c$ the speed of light in vacuum and $v_g$ the group velocity of light, and $h$ the thickness of the membrane. Our calculations show that the group delay reaches up to 5.42 ns, and $n_g$ can be as high as 1.63×10$^6$. This indicates that the group velocity of light propagating through the free-standing cSi-membrane metasurface is 1.63×10$^6$ times slower than that in vacuum, confirming the strong slow light effect due to the q-BIC-EIT resonance (Fig. S3).

A major strength of our system is its ability to tune the location of the resonance peak as well as its Q-factor, thereby allowing for a compact and robust optical sensor realization. This adaptability and scalability facilitate the integration into optical devices without compromising size or system simplicity. To demonstrate the spectral tunability of the q-BIC-EIT resonances, we introduced a scaling factor, *S*, which modifies the dimensions of the unit cell in Figure 3a while keeping h = 1 µm and θ = 10°. Figures 3g and 3h show the experimental and simulated transmittance spectra obtained by varying S from 0.7 to 1.0. The shape of the measured transmittance peak shows good agreement with the simulation, although the Q-factor is reduced. This is most likely due to imperfections in the fabricated sample, introducing scattering loss and disrupting coherence among the resonators.

To investigate the effect of the tilting angle ($\theta$) on the q-BIC-EIT resonances, we fabricated an array of nine q-BIC-EIT metasurfaces each with a 400 µm x 400 µm area on a single free-standing Si-membrane (see Figure 4a). While keeping all other geometrical parameters of the unit cell constant, we varied $\theta$ between 0° and 8°. At $\theta$ = 0°, the transparency window vanishes ($\Gamma_{BIC} \rightarrow 0$ and the BIC Q-factor $\rightarrow \infty$), leaving only the SLM resonance (dashed line in Figure 4b) with the Q-factor of the q-BIC-EIT mode approaching zero. Among the $\theta \neq 0°$ designs, the Q-factor increases dramatically as the tilting angle decreases ($\theta \rightarrow 0$), with the highest measured value being approximately 734 (Fig. 4b inset). The experimentally extracted Q-factors align well with those obtained from simulations, confirming the q-BIC nature of the resonances (Q-factor $\propto \theta^{-2}$ Fig. 4c). It is also noteworthy that in the simulations, the transmittance peak amplitude remains unaffected by decrease in the tilting angle $\theta$ (see Figure 4d). The apparent reduction in peak amplitude observed in our experimental measurements (Figure 4b) is likely due to the spectral resolution limit of our optical setup and the finite size of the metasurface patterned area.

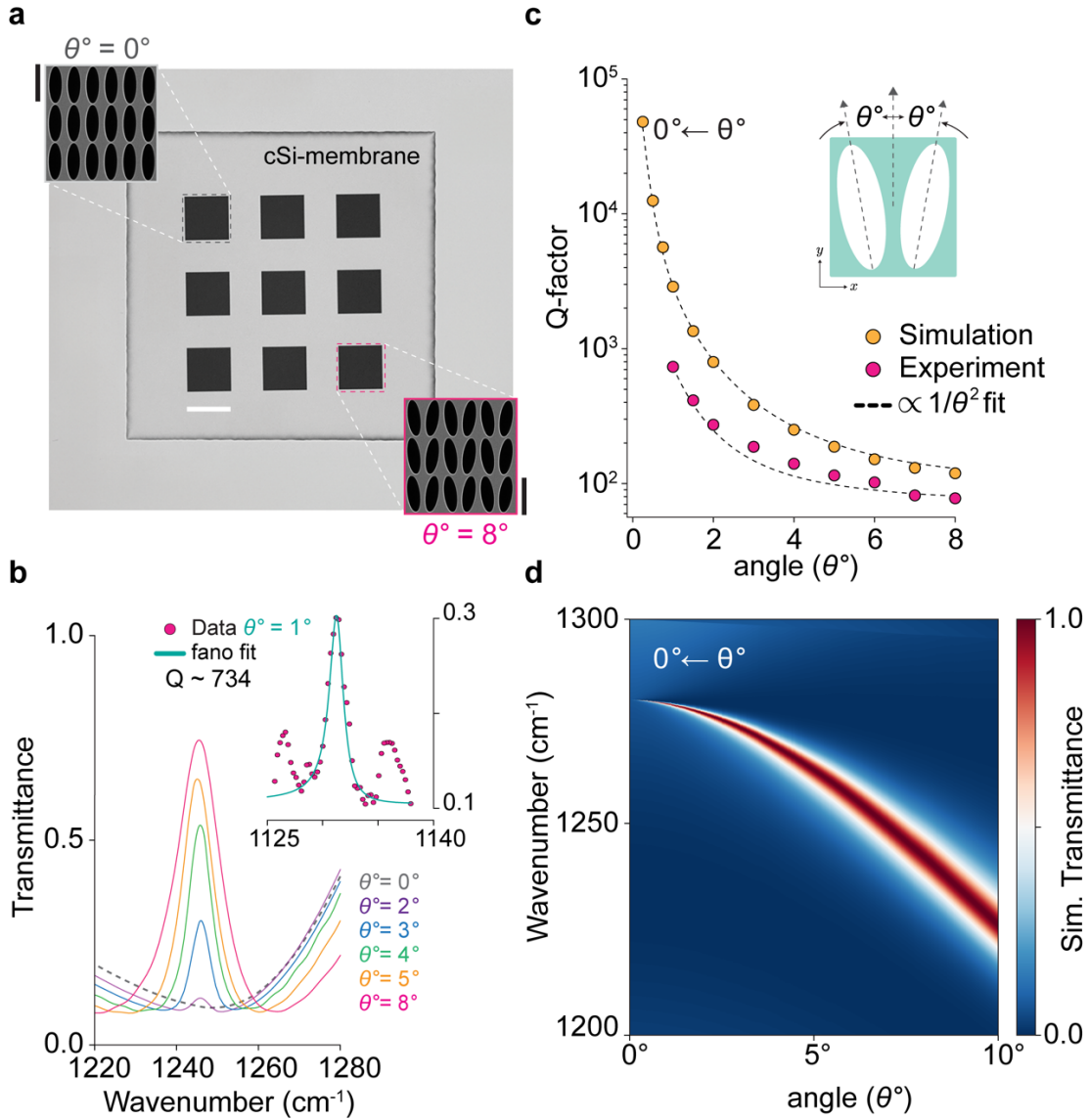

**Figure 4. Tuning of the resonance Q factor via the angle variation. a,** Brightfield photograph and SEM images (insets; scale bar 4 $\mu$m) of nine metasurfaces (each 400 $\mu$m by 400 $\mu$m) with $\theta$ varying from 0° to 8° degrees. **b**, As the tilting angle approaches zero the measured q-BIC-EIT resonances become sharper, and the Q factor increases. The highest measured Q factor is ~734 for $\theta = 1°$ (shown in the inset). **c**, Evolution of Q-factor as a function of asymmetry parameter $\theta$ for simulated and experimental data (Q-factor $\propto \theta^{-2}$). **d**, Simulated transmittance spectra of a metasurface with S = 1.02 and a mid-infrared illumination source polarized along the x-axis with $\theta$ varying from 0° to 10°. As $\theta$ approaches 0°, the Q factor approaches infinity, and the resonance linewidth dramatically decreases.

## Vibrational strong coupling (VSC) with EIT-q-BIC

To investigate the near-field light-matter interactions in our metasurface photonic cavities, we coated the patterned free-standing membranes with PMMA layers of three different thicknesses—30 nm, 60 nm, and 110 nm (Fig. 5a and Methods). Figure 5b shows the experimental and simulated spectral maps obtained by coating the free-standing membrane metasurface arrays with a 110 nm PMMA layer while varying the metasurface scaling factor (S) from 0.72 to 0.86. When the frequency of the swept q-BIC-EIT resonance mode overlaps with the C=O vibrational band of PMMA (~ 1730 cm$^{-1}$), a splitting of the q-BIC-EIT peak is observed in the transmission spectrum. This splitting corresponds to two polariton branches of different energies; the higher energy state is known as the upper polariton ($\hbar\omega_{UP}$), and the lower energy state is the lower polariton ($\hbar\omega_{LP}$) (26, 27). Consequently, the polariton formation reveals the typical anti-crossing pattern characteristic of strongly coupled systems. In the simulation, the free-standing tilted rod metasurface was coated with a 110 nm-thick PMMA, as illustrated in Figure 5a, showing excellent agreement with the experimental results.

In Figure 5c, we report the measured spectra of the bare metasurfaces (cyan) and the measured PMMA reference signal (grey) for 30, 60, and 110 nm-thick PMMA spin-coated on unpatterned Si-membranes. The full-width-half-maximum (FWHM$_{PMMA}$ ~ 26.8 cm$^{-1}$) of the PMMA's C=O vibrational band ~ 1730 cm$^{-1}$ is highlighted with the grey-shaded region. The absorbance of the PMMA C=O bond on unpatterned membranes cannot be detected for the 30 nm and 60 nm PMMA layers because these signals are below the noise limit of our measurement system, while the 110 nm PMMA layer generates a weak absorption signal. In contrast, when the similar thicknesses of PMMA layers are measured by coating them on the metasurface, a Rabi mode splitting and polariton formation is observed due to vibrational strong coupling between the metasurface resonance and the vibrational mode of PMMA bond (see Figure 5d and e). This drastic difference between the absorbance fingerprint of the PMMA bond on the unpatterned membrane vs. the metasurface can be attributed to the coherent interaction of PMMA molecules with the localized electromagnetic fields in the tilted rod cavities. Furthermore, in Figure 5f, we show infrared transmission microscopy images of a metasurface coated with 60 nm thick PMMA to reveal the transmittance contrast in the 2D space captured at the upper and lower polariton peak wavenumbers and the PMMA absorption band.

In coupled cavity-molecule systems, it is a common practice to investigate the coupling regime to verify if the ensemble fulfills the strong coupling criteria, which is satisfied when the frequency separation of the upper and lower polariton peaks, also known as Rabi splitting ($\Omega$) is larger than the sum of the loss rate of the PMMA absorption band ($\Gamma_{PMMA}$) and the cavity mode ($\Gamma_{q-BIC-EIT}$, $\theta° = 10°$). This criterion is mathematically expressed as: $\Omega_{exp} > \Gamma_{PMMA} + \Gamma_{q-BIC-EIT}$. For the three distinct PMMA thicknesses of 30 nm, 60 nm, and 110 nm, we measured Rabi splitting as 32, 36, and 48 cm$^{-1}$, respectively, at zero detuning, when the frequency of the q-BIC-EIT mode perfectly overlaps with the C=O vibrational bond in PMMA ($\delta = \omega_{PMMA} - \omega_{res} = 0$). All three measurements satisfied the strong coupling criterion of ($\Omega_{exp} > \Gamma_{PMMA} + \Gamma_{q-BIC-EIT} = 13.4 + 15.3 = 28.7$ cm$^{-1}$). Supplementary Video 1 illustrates this unique resonant behavior, where each frame represents the transmittance response at a specific wavenumber for a q-BIC-EIT metasurface (400 μm × 400 μm), with and without a ~110 nm PMMA thin layer.

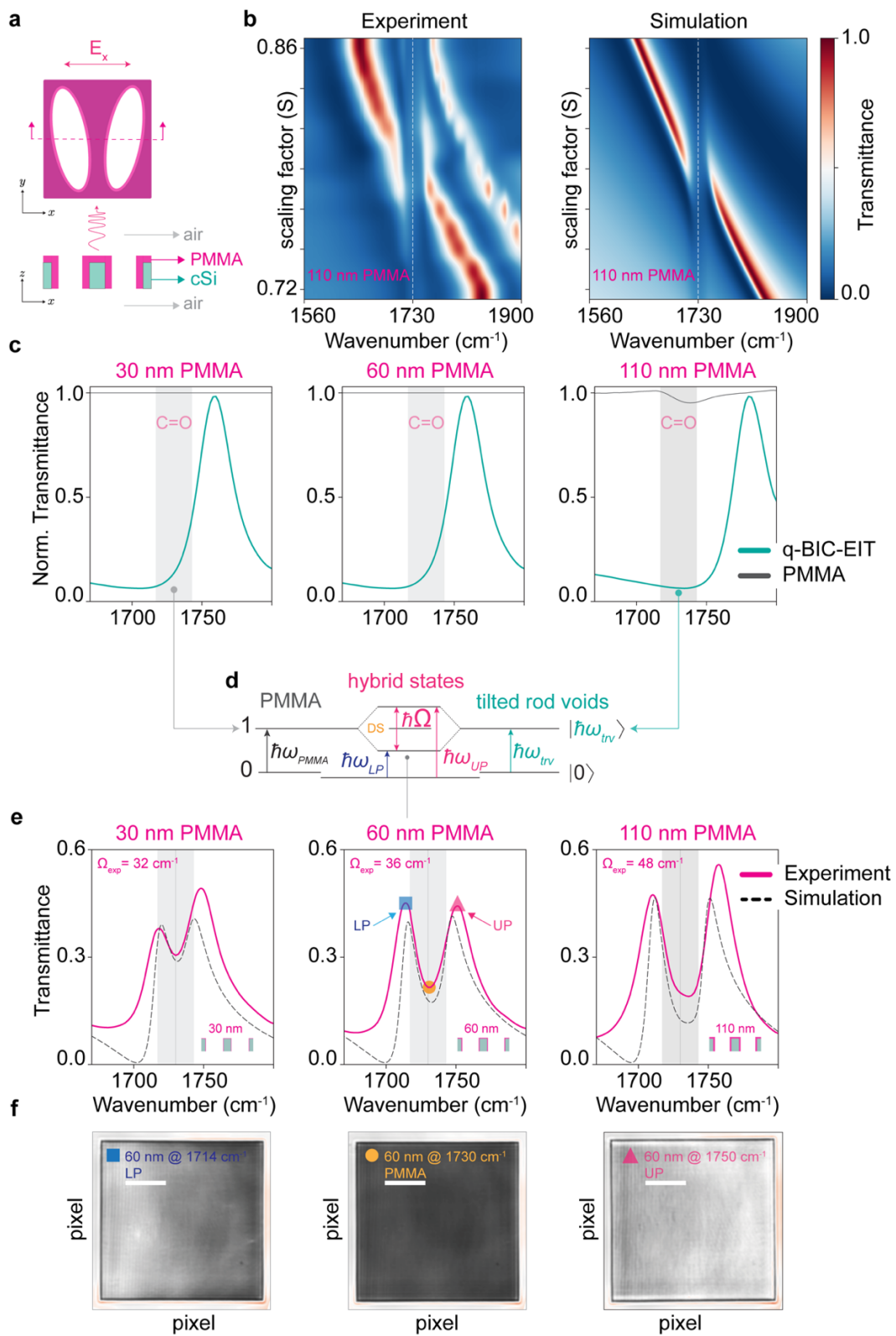

**Figure 5. Vibrational strong coupling (VSC) between the q-BIC-EIT-like resonances and vibrational transitions of PMMA molecules**. **a**, Illustration of a unit cell with tilted rod voids when coated with a thin PMMA layer, showing top and cross section views. **b**, Experimentally measured and simulated spectral maps generated by sweeping the resonance peak through PMMA's C=O absorption band at 1730 cm$^{-1}$. Both spectral maps show anti-crossing patterns due to vibrational strong coupling between a 110 nm PMMA layer and metasurface resonance. **c**, Experiment response of thin layers of PMMA (30 nm, 60 nm, and 110 nm) spin-coated on the top face of an unpatterned cSi membranes (grey lines), and bare metasurface q-BIC-EIT-like mode resonances (cyan lines). **d**, Strong coupling between the PMMA molecule's vibrational transition and the metasurface resonance modes leads to the formation of polaritonic states, $\hbar\omega_{UP}$ and $\hbar\omega_{LP}$, which are separated in energy by the Rabi splitting $\hbar\Omega$, and the uncoupled PMMA molecules populate the dark states (DS). **e**, Experimentally measured and simulated spectra showing polariton formation when metasurfaces are coated with three distinct PMMA thickness 30 nm, 60 nm, and 110 nm. When the metasurface resonance frequency matches the vibrational mode of the PMMA C=O bond around 1730 cm$^{-1}$ (dashed line) the system degenerates into two hybrid states; the higher energy state ($\hbar\omega_{UP}$) is known as the upper polariton (UP), and the lower energy state ($\hbar\omega_{LP}$) is the lower polariton (LP). **f**, Infrared transmission microscopy images of the metasurface (400 $\mu$m × 400 $\mu$m) spin-coated with 60 nm of PMMA under VSC regime at 1714 cm$^{-1}$ (LP), 1730 cm$^{-1}$ (PMMA C=O band) and at 1750 cm$^{-1}$ (UP) (scale bar 100 $\mu$m).

**Sensitive protein monolayer detection**

Finally, we demonstrate how our free-standing membrane metasurfaces supporting EIT-like resonance peaks in transmission mode can be used for sensitive biomolecule detection. To achieve this, we designed and fabricated an array of 18 metasurfaces (400 μm × 400 μm), each tuned using the S parameter to cover the characteristic protein fingerprints from 1500 to 1750 cm$^{-1}$, specifically targeting the amide I (~1656 cm$^{-1}$) and amide II (~1542 cm$^{-1}$) bands. Next, we coated the metasurface arrays with a monolayer of protein A/G using the physical absorption method (Fig. 6a and Methods). Figure 6b shows measured and simulated spectra (solid lines) acquired from arrays of protein-coated metasurfaces after normalized to their individual resonance peak references acquired before protein coating. The reference signal shown as a dashed line in Figure 6b was measured by drop-casting a high-concentration protein A/G solution, letting it dry to form a thick layer of proteins on a standard CaF$_2$ substrate, and measuring it in transmission mode. Our findings show that the enhanced light-matter interactions between the metasurface cavities and a monolayer of proteins cause a significant dampening of the q-BIC-EIT resonance peaks in transmission mode. The transmittance amplitude changes in individual metasurface spectra correlate with the amide I (~1656 cm$^{-1}$) and amide II (~1542 cm$^{-1}$) bands, capturing proteins' characteristic spectral fingerprints.

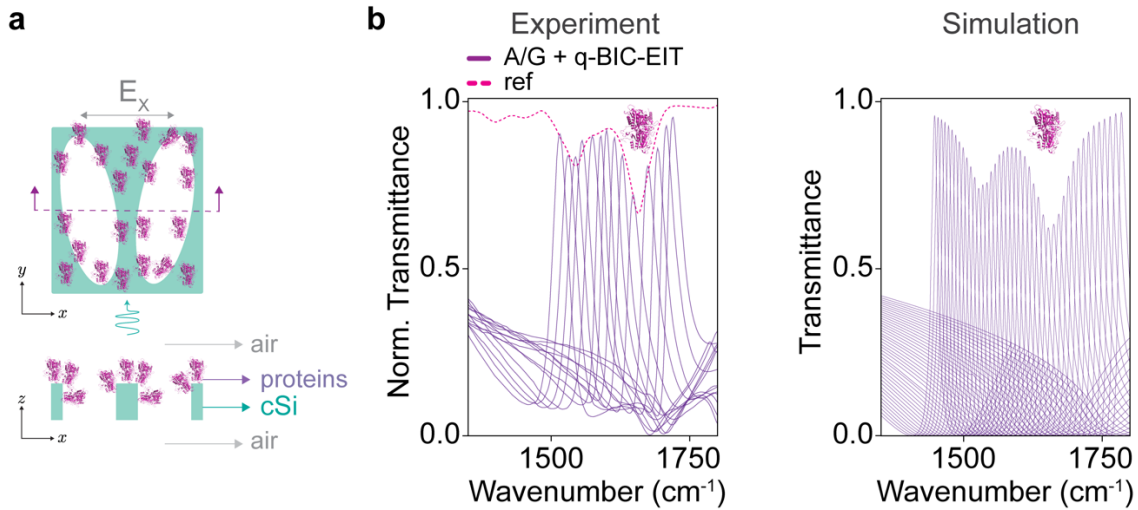

**Figure 6. Protein monolayer detection with q-BIC-EIT-like resonance modes. a**, Illustration of a monolayer of protein A/G coating on the metasurface. **b**, When coupled to the q-BIC-EIT-like resonance modes, the absorption signals associated with the protein bands, amide I and II, are enhanced and become detectable in transmission mode. The dashed line shows a reference protein absorption signal collected by drop-casting a bulk A/G protein solution onto a $CaF_2$ substrate. The resonance curves were collected from 18 metasurfaces (400 μm × 400 μm), each tuned using the S parameter to resonate at the amide I and amide II protein bands. In the simulation, we modeled a protein monolayer with a 10 nm continuous film covering the top surface and inner walls of the metasurface. Simulation results shown on the right are in good agreement with the experimental measurements.

## Discussion

One of the primary challenges in the development of high-Q metasurfaces for biochemical sensing has been their integration into compact and robust optical systems. So far, metasurface devices supporting high-Q resonances, such as qBIC modes, surface lattice modes, guided modes, etc., show a peak in reflection mode, which complicates their use in simple and compact imaging systems. To overcome this limitation, we demonstrated a new type of metasurface, which supports high-Q transmissive resonance modes for biochemical sensing applications in the mid-IR spectral range. By engraving a carefully designed periodic air cavity pattern into a free-standing Si membrane, we engineered a subwavelength thick (1 μm < $\lambda_{mid-IR}$~ 5-10 μm) media, where the qBIC modes spectrally and spatially overlap with SLMs. The destructive interference of these two nonradiative modes within the same media induces a narrow (Q~734 @ $\lambda$~8.8 μm) transparency window, generating an analogue of the EIT modes. Notably, on the same Si membrane, the qBIC-EIT modes can be geometrically tuned over a wide spectral range (1000–1800 $cm^{-1}$), which overlaps with the characteristic vibrational fingerprints of many organic compounds. Moreover, by varying the asymmetry in the geometry of the metaunit elements, we showed the regulation of the resonance width. In sum, our new photonic device approach enables customizable spectral and temporal light localization and facilitates the integration into collinear optical paths, providing a versatile platform for broad mid-IR biochemical sensing applications.

The results of this study will open several avenues for future research, spurring from the applications we showed in both strong and weak coupling regimes of cavity-coupled systems. The ability to achieve vibrational strong coupling with a thinly coated PMMA film, characterized by significant Rabi splitting (32 cm$^{-1}$), underscores the potential of these metasurfaces for various polaritonic chemistry applications. The sharp resonances in the mid-IR are highly sensitive to their near-field media, as evidenced by our successful biosensing of protein monolayers in transmission mode. Thus, the proposed metasurfaces can catalyze the detection of fine variations in the spectral fingerprints of functional biomolecules, which has implications for biomarker detection in medical diagnostics. Furthermore, our metasurface design localizes the electromagnetic field into accessible air cavities, which are well suited for real-time gas and liquid sensing when integrated into flow-through chambers for environmental monitoring and industrial process control applications. Notably, our Si-membrane metasurfaces are fully compatible with mass-scale chip manufacturing foundries, which is key to its technology translation into products. In conclusion, this work not only advances the fundamental understanding of metasurface physics but also holds promise for practical applications in various fields requiring precise and reliable biochemical detection.

## MATERIALS AND METHODS

**Numerical simulations**

The finite-element frequency-domain solver (CST Microwave Studio 2023, Dassault Systèmes, France) was used to calculate the transmittance spectra. The electric field enhancement in Figure 3b, 3c, and 3d was calculated using Tidy3D FDTD (Flexcompute, California, USA). Similarly, the scattering cross-section, $\sigma_{scs}$, of cartesian multipoles was computed using Lumerical FDTD, plus an open-source Matlab implementation of multipole expansion for applications in nanophotonics. In all simulations, periodic boundary conditions were used along the x- and y-axes and perfectly matched layers (PML) along the z-direction. We assumed plane wave illumination with a Gaussian temporal profile from the top of the structure at normal incidence. The optical constants for the cSi were taken from(*28*) and for PMMA we used the PMMA refractive index from(*29*). The cavity effective mode volume was calculated as(*31*):

$$V_{eff} = \frac{\int \varepsilon(\boldsymbol{r})|\boldsymbol{E}(\boldsymbol{r})|^2 d^3\boldsymbol{r}}{\max{(\varepsilon(\boldsymbol{r})|\boldsymbol{E}(\boldsymbol{r})|^2)}}$$

**Device fabrication**

Intrinsic single-crystal 1-μm thick silicon membranes (2.8 mm × 2.8 mm), supported by a 300 μm-thick silicon frame (10 mm × 10 mm), were used as substrates to fabricate the metasurfaces (Norcada, Alberta, Canada) with a high fabrication yield of about 95%. Free-standing tilted rod metasurfaces were fabricated as follows: electron-beam lithography, lift-off, and reactive ion silicon etching. Initially, a ~50 nm thin layer of SiO$_2$ hard mask was deposited on the bare Si-membrane using Plasma-Enhanced Chemical Vapor Deposition (PECVD) at 100°C, 2.5 mT, 1200 W inductive coupling plasma (ICP), 20 sccm N$_2$O, and 8.5 sccm SiH$_4$. We carefully positioned the cSi-membrane at the center of a 20 mm × 20 mm silicon chip, which served as a sample carrier. The corners of the membrane were taped using Kapton tape to the carrier to avoid damaging the cSi-membrane during the spin-coating process. This was followed by spin-coating ~110 nm of ZEP resist (1:1) at 4000 rpm and baking the cSi-membrane for 3 minutes at 160°C. Metasurfaces of 400 μm × 400 μm were then patterned using a 100 keV electron beam (JEOL JBX-8100FS, 4 nA and a dose of 250 μC/cm²). After the resist exposure, the ZEP on the Si membranes was developed in n-Amyl acetate for 60 seconds, followed by soaking in IPA for 60 seconds. Next, the hole shapes were etched in the thin $SiO_2$ layer using reactive ion etching at 20°C, 10 mT, 50 W RIE, 50 sccm $CHF_3$, 2 sccm $O_2$, and 10 T He backside pressure. The residual ZEP was stripped off using oxygen plasma (170 mT, 24 sccm $O_2$, 150 W for 3 minutes). The geometry was transferred from the $SiO_2$ into the 1 μm-thick Si-membrane using reactive ion etching with HBr (20°C, 10 mT, 20 sccm $Cl_2$, 300 W RIE, 1000 W ICP, 10 T He backside pressure) for 6 seconds and then (20°C, 12 mT, 50 sccm HBr, 2 sccm $O_2$, 100 W RIE, 250 W ICP, 10 T He backside pressure) for 8 minutes. Finally, the residual $SiO_2$ was removed using reactive ion etching.

**Cartesian Multipoles analysis**

We computed the total scattering cross section ($\sigma_{scs}$) spectra of the cartesian multipoles (*32*) in free space as

$$\sigma_{scs} = \frac{k^4}{6\pi\varepsilon_o^2|E_o|^2}\left[\sum_{\alpha,\beta}(|p_\alpha|^2 + \left|\frac{m_\alpha}{c}\right|^2) + \frac{1}{120}\sum_{\alpha,\beta}(\left|\widehat{Q^e}_{\alpha,\beta}\right|^2 + \left|\frac{\widehat{Q^m}_{\alpha,\beta}}{c}\right|^2)\right]$$

Where $k$, $\varepsilon_o$, an $c$ refers to the wavenumber, electric permittivity of free-space, and speed of light in free-space, respectively, $\alpha, \beta = (x, y, z)$, and $E_o$ stands for the incident electric field. The terms

$$p_\alpha = -\frac{1}{i\omega}\left[\int J_\alpha j_o(k\,r^3)d^3r + \frac{k^2}{2}\int(3(r\cdot J)r_\alpha - r^2 J_\alpha)\frac{j_2(kr)}{(kr)^2}d^3r\right]$$

$$m_\alpha = \frac{3}{2}\int(r\times J)_\alpha\frac{j_1(kr)}{kr}d^3r$$

$$\widehat{Q^e}_{\alpha,\beta} = -\frac{3}{i\omega}\left[\int(3(r_\beta J_\alpha + r_\alpha J_\beta) - 2(r\cdot J)\delta_{\alpha\beta})\frac{j_1(kr)}{kr}d^3r\right.$$
$$\left. + 2k^2\int(5\,r_\alpha r_\beta\,(r\cdot J) - r^2(r_\alpha J_\beta + r_\beta J_\alpha) - r^2(r\cdot J)\delta_{\alpha\beta})\frac{j_3(kr)}{(kr)^3}d^3r\right.$$
$$\widehat{Q^m}_{\alpha,\beta} = 15\int(r_\alpha(r\times J)_\beta + r_\beta(r\times J)_\alpha)\frac{j_2(kr)}{(kr)^2}d^3r$$

Are the moments of the electric dipole (ED, $p_\alpha$), magnetic dipole (MD, $m_\alpha$), electric quadrupole (EQ, $\widehat{Q^e}_{\alpha,\beta}$), and magnetic quadrupole (MQ, $\widehat{Q^m}_{\alpha,\beta}$), respectively. Here, $j_n(\rho)$ denotes the spherical Bessel function defined by $j_n(\rho) = \sqrt{\pi/2\rho}B_{n+1/2}(\rho)$, where $B_{n+1/2}(\rho)$ is the Bessel function of first kind.

When incident light excites the Si-membranes, the induced current density distributions $J(r)$ can be obtained from the electric field distributions $E(r)$ by

$$J(r) = -i\omega\varepsilon_\circ(n^2 - 1)E(r)$$

With $n$ the refractive index of the resonator.

**PMMA spin coating and thickness characterization**

All PMMA spin-coating was applied only to the top side of the cSi-metasurface. Various thicknesses of PMMA (30 nm, 60 nm, and 110 nm) were achieved by using PMMA 950 A2 (Kayaku Advanced Material, Massachusetts, USA) with different spin-coating speeds, and dilutions in anisole. Finally, the metasurface was baked for 5 minutes at 180°C. To measure the PMMA thickness, each metasurface coated with PMMA was accompanied by a bare Si chip of the same size, also coated with PMMA using identical parameters. We used an ellipsometer (J.A. Woollam, Nebraska, USA) and a reflectometer (Filmetrics, California, USA) to measure the PMMA thickness on the bare Si substrates. The full-width at half-maximum (FWHM ~ 26.8 cm$^{-1}$ → $\Gamma_{PMMA} = 13.4$ cm$^{-1}$) of the 1730 cm$^{-1}$ vibrational mode of PMMA was measured by fitting a Gaussian function to the spectrum of 110 nm PMMA spin-coated on an unpatterned Si membrane.

**Protein coating**

For the protein sensing measurements, protein A/G was diluted in 10 mM acetate solution at 0.5 mg/mL concentration. The top side of the tilted rod free-standing metasurface was incubated with this protein A/G solution for 2 hours to allow protein physisorption. After incubation, the metasurface was rinsed with deionized water to remove unbound protein and agglomerates, and then dried with nitrogen.

**Optical setup and measurements**

Mid-infrared spectral measurements were done, unless otherwise noted, using a tunable quantum cascade laser (QCL) integrated into a mid-infrared microscope (Spero-QT, Daylight Solutions, California, USA). Using four QCL modules, the microscope can collect spectra covering the fingerprint spectral region from 950 to 1800 cm$^{-1}$ with 2 cm$^{-1}$ spectral resolution. During acquisition, the sample chamber was continuously purged with dry nitrogen to clear out water vapor. The free-standing metasurface was illuminated with collimated, linearly polarized, light at normal incidence; the spectral data was acquired in transmission mode using a 12.5 × IR collection objective (0.7 NA) and detected using an uncooled microbolometer focal plane array with 480 × 480 pixels obtaining a field of view of 650 μm × 650 μm. In transmission mode, our chemical microscope captures the sample's decadic absorbance as: A = -log$_{10}$(T), where T = I/I$_0$ and I$_0$ is the transmission through unpatterned Si-membrane as a sample. From the definition of decadic absorbance, we then calculate the transmittance of the sample as follows: T =10$^{-A}$.

Figures 4b, 4c, and 5b were composed using data collected by a Fourier-transform infrared (FTIR) spectrometer, coupled to an infrared microscope (Bruker Vertex 70 FTIR and Hyperion 2000). Transmittance spectra were collected using linearly polarized light through a low NA refractive

objective (5×, 0.17 NA, Pike Technology, Wisconsin, USA). The setup employed collimated light, achieved by removing the bottom condenser, and a liquid-nitrogen-cooled mercury-cadmium-telluride (MCT) detector to measure the spectra. All acquired transmittance spectra of the cSi-membrane metasurfaces were normalized using the transmission spectrum of an unpatterned Si-membrane.

**Data processing**

The measured transmittance spectrum in Figures 3g, 5b, and 5c was normalized to its maximum for visualization purposes. The experimental transmittance in Figures 5e and 6b (experimental) was normalized to the bare tilted rod metasurface spectral response using Matlab's control point image registration routine and in-house scripts written in Python. All cavity losses and Q-factors were obtained by fitting the raw transmittance data shown in Figure 4b with the following Fano fit from (*33*):

$$T = \left| ie^{i\phi} t_0 + \frac{\Gamma_R}{\Gamma_R + \Gamma_{NR} + i(\lambda - \lambda_{res})} \right|^2$$

Where $ie^{i\phi} t_0$ describes the background and the shape of the resonance. The Q-factor can then be calculated as:

$$Q = \frac{\lambda_{res}}{2(\Gamma_R + \Gamma_{NR})}$$

Where $\lambda_{res}$, $\Gamma_R$, and $\Gamma_{NR}$ are the resonance wavelength, radiative, and absorptive loss rates, respectively.

**Data availability**

The data that support the findings of this study are available from the corresponding author upon reasonable request.

**Competing Interest**

The authors declare no competing interest.

**Acknowledgement**

The authors thank Professor Eduardo R. Arvelo for his assistance with rendering and illustration in Blender, and Professor Jennifer Choy for providing access to the FDTD computation tool in her lab. Both professors are affiliated with the Electrical & Computer Engineering Department at UW-Madison. The authors also thank PhD student Justin Edwards for facilitating a macro lens to take close-up photographs of the Bucky-metasurface. The authors gratefully acknowledge use of facilities and instrumentation in the UW-Madison Wisconsin Center for Nanoscale Technology. The Center (wcnt.wisc.edu) is partially supported by the Wisconsin Materials Research Science and Engineering Center (NSF DMR-2309000) and the University of Wisconsin-Madison. Photonic metasurfaces were fabricated at the Center for Nanoscale Materials at the Argonne National Laboratory, a U.S. Department of Energy Office of Science User Facility, was supported by the U.S. DOE, Office of Basic Energy Sciences, under Contract No. DE-AC02-06CH11357. F.Y. acknowledges financial support from the U.S. National Science Foundation (grant no. 2401616) and the U.S. National Institutes of Health (grant no. R21EB034411).

# References

1. A. I. Kuznetsov, M. L. Brongersma, J. Yao, M. K. Chen, U. Levy, D. P. Tsai, N. I. Zheludev, A. Faraon, A. Arbabi, N. Yu, D. Chanda, K. B. Crozier, A. V. Kildishev, H. Wang, J. K. W. Yang, J. G. Valentine, P. Genevet, J. A. Fan, O. D. Miller, A. Majumdar, J. E. Fröch, D. Brady, F. Heide, A. Veeraraghavan, N. Engheta, A. Alù, A. Polman, H. A. Atwater, P. Thureja, R. Paniagua-Dominguez, S. T. Ha, A. I. Barreda, J. A. Schuller, I. Staude, G. Grinblat, Y. Kivshar, S. Peana, S. F. Yelin, A. Senichev, V. M. Shalaev, S. Saha, A. Boltasseva, J. Rho, D. K. Oh, J. Kim, J. Park, R. Devlin, R. A. Pala, Roadmap for Optical Metasurfaces. *ACS Photonics* **11**, 816–865 (2024).

2. A. Krasnok, M. Caldarola, N. Bonod, A. Alú, Spectroscopy and Biosensing with Optically Resonant Dielectric Nanostructures. *Advanced Optical Materials* **6**, 1701094 (2018).

3. J. Hu, F. Safir, K. Chang, S. Dagli, H. B. Balch, J. M. Abendroth, J. Dixon, P. Moradifar, V. Dolia, M. K. Sahoo, B. A. Pinsky, S. S. Jeffrey, M. Lawrence, J. A. Dionne, Rapid genetic screening with high quality factor metasurfaces. *Nat Commun* **14**, 4486 (2023).

4. S. K. Biswas, W. Adi, A. Beisenova, S. Rosas, E. R. Arvelo, F. Yesilkoy, From weak to strong coupling: quasi-BIC metasurfaces for mid-infrared light–matter interactions. *Nanophotonics* **13**, 2937–2949 (2024).

5. A. Kodigala, T. Lepetit, Q. Gu, B. Bahari, Y. Fainman, B. Kanté, Lasing action from photonic bound states in continuum. *Nature* **541**, 196–199 (2017).

6. H. Qin, Y. Shi, Z. Su, G. Wei, Z. Wang, X. Cheng, A. Q. Liu, P. Genevet, Q. Song, Exploiting extraordinary topological optical forces at bound states in the continuum. *Science Advances* **8**, eade7556 (2022).

7. S. Yang, C. Hong, Y. Jiang, J. C. Ndukaife, Nanoparticle Trapping in a Quasi-BIC System. *ACS Photonics* **8**, 1961–1971 (2021).

8. A. S. Solntsev, G. S. Agarwal, Y. S. Kivshar, Metasurfaces for quantum photonics. *Nat. Photonics* **15**, 327–336 (2021).

9. K. Koshelev, S. Kruk, E. Melik-Gaykazyan, J.-H. Choi, A. Bogdanov, H.-G. Park, Y. Kivshar, Subwavelength dielectric resonators for nonlinear nanophotonics. *Science* **367**, 288–292 (2020).

10. A. Tittl, A. Leitis, M. Liu, F. Yesilkoy, D.-Y. Choi, D. N. Neshev, Y. S. Kivshar, H. Altug, Imaging-based molecular barcoding with pixelated dielectric metasurfaces. *Science* **360**, 1105–1109 (2018).

11. M. Minkov, D. Gerace, S. Fan, Doubly resonant $\chi^{(2)}$ nonlinear photonic crystal cavity based on a bound state in the continuum. *Optica, OPTICA* **6**, 1039–1045 (2019).

12. K. Sun, M. Sun, Y. Cai, U. Levy, Z. Han, Strong coupling between quasi-bound states in the continuum and molecular vibrations in the mid-infrared. *Nanophotonics* **11**, 4221–4229 (2022).

13. W. Adi, S. Rosas, A. Beisenova, S. K. Biswas, F. Yesilkoy, Trapping light in air with membrane metasurfaces for vibrational strong coupling, *arXiv.org* (2024). https://arxiv.org/abs/2402.17901v1.

14. A. Leitis, A. Tittl, M. Liu, B. H. Lee, M. B. Gu, Y. S. Kivshar, H. Altug, Angle-multiplexed all-dielectric metasurfaces for broadband molecular fingerprint retrieval. *Science Advances* **5**, eaaw2871 (2019).

15. A. Leitis, M. L. Tseng, A. John-Herpin, Y. S. Kivshar, H. Altug, Wafer-Scale Functional Metasurfaces for Mid-Infrared Photonics and Biosensing. *Advanced Materials* **33**, 2102232 (2021).

16. L. Nan, A. Mancini, T. Weber, G. L. Seah, E. Cortés, A. Tittl, S. A. Maier, Highly confined incident-angle-robust surface phonon polariton bound states in the continuum metasurfaces. arXiv arXiv:2403.18743 [Preprint] (2024). http://arxiv.org/abs/2403.18743.

17. A. John-Herpin, A. Tittl, L. Kühner, F. Richter, S. H. Huang, G. Shvets, S.-H. Oh, H. Altug, Metasurface-Enhanced Infrared Spectroscopy: An Abundance of Materials and Functionalities. *Advanced Materials* **35**, 2110163 (2023).

18. Y. Yang, I. I. Kravchenko, D. P. Briggs, J. Valentine, All-dielectric metasurface analogue of electromagnetically induced transparency. *Nat Commun* **5**, 5753 (2014).

19. C. Wang, X. Jiang, W. R. Sweeney, C. W. Hsu, Y. Liu, G. Zhao, B. Peng, M. Zhang, L. Jiang, A. D. Stone, L. Yang, Induced transparency by interference or polarization. *Proc. Natl. Acad. Sci. U.S.A.* **118**, e2012982118 (2021).

20. N. Liu, L. Langguth, T. Weiss, J. Kästel, M. Fleischhauer, T. Pfau, H. Giessen, Plasmonic analogue of electromagnetically induced transparency at the Drude damping limit. *Nature Mater* **8**, 758–762 (2009).

21. X. Zhao, R. Huang, X. Du, Z. Zhang, G. Li, Ultrahigh-Q Metasurface Transparency Band Induced by Collective–Collective Coupling. *Nano Lett.* **24**, 1238–1245 (2024).

22. G. W. Castellanos, P. Bai, J. Gómez Rivas, Lattice resonances in dielectric metasurfaces. *Journal of Applied Physics* **125**, 213105 (2019).

23. K. Koshelev, S. Lepeshov, M. Liu, A. Bogdanov, Y. Kivshar, Asymmetric Metasurfaces with High-$Q$ Resonances Governed by Bound States in the Continuum. *Phys. Rev. Lett.* **121**, 193903 (2018).

24. A. Chukhrov, S. Krasikov, A. Yulin, A. Bogdanov, Excitation of a bound state in the continuum via spontaneous symmetry breaking. *Phys. Rev. B* **103**, 214312 (2021).

25. B. Peng, Ş. K. Özdemir, W. Chen, F. Nori, L. Yang, What is and what is not electromagnetically induced transparency in whispering-gallery microcavities. *Nat Commun* **5**, 5082 (2014).

26. T. W. Ebbesen, A. Rubio, G. D. Scholes, Introduction: Polaritonic Chemistry. *Chem. Rev.* **123**, 12037–12038 (2023).

27. F. J. Garcia-Vidal, C. Ciuti, T. W. Ebbesen, Manipulating matter by strong coupling to vacuum fields. *Science* **373**, eabd0336 (2021).

28. H. H. Li, Refractive index of silicon and germanium and its wavelength and temperature derivatives. *Journal of Physical and Chemical Reference Data* **9**, 561–658 (1980).

29. S. Tsuda, S. Yamaguchi, Y. Kanamori, H. Yugami, Spectral and angular shaping of infrared radiation in a polymer resonator with molecular vibrational modes. *Opt. Express, OE* **26**, 6899–6915 (2018).

31. M. Boroditsky, R. Coccioli, E. Yablonovitch, Y. Rahmat-Samii, K. W. Kim, Smallest possible electromagnetic mode volume in a dielectric cavity. *IEE Proceedings - Optoelectronics* **145**, 391–397 (1998).

32. T. Hinamoto, M. Fujii, MENP: an open-source MATLAB implementation of multipole expansion for nanophotonics. *OSA Continuum* **4**, 1640 (2021).

33. S. Fan, W. Suh, J. D. Joannopoulos, Temporal coupled-mode theory for the Fano resonance in optical resonators. *Journal of the Optical Society of America A* **20**, 569 (2003).